# A Mathematical Model for Predicting the Life of PEM Fuel Cell Membranes Subjected to Hydration Cycling

S.F. Burlatsky[1], M. Gummalla[1*], J. O'Neill[4],

V.V. Atrazhev[2,3], A.N. Varyukhin[3], D.V. Dmitriev[2,3], N.S. Erikhman[2,3]

[1] United Technologies Research Center, 411 Silver Lane, East Hartford, CT 06108, USA

[2] Russian Academy of Science, Institute of Biochemical Physics, Kosygin str. 4, Moscow, 119334, Russia

[3] Science for Technology LLC, Leninskiy pr-t 95, 119313, Moscow, Russia

[4] UTC Power, 195 Governor's Highway, South Windsor, CT 06074, USA

*Corresponding Author, Phone: 1 860-610-7383, Fax: 1 860-660-8240, Email: gummalm@utrc.utc.com

## Abstract

Under typical PEM fuel cell operating conditions, part of membrane electrode assembly is subjected to humidity cycling due to variation of inlet gas RH and/or flow rate. Cyclic membrane hydration/dehydration would cause cyclic swelling/shrinking of the unconstrained membrane. In a constrained membrane, it causes cyclic stress resulting in mechanical failure in the area adjacent to the gas inlet. A mathematical modeling framework for prediction of the lifetime of a PEM FC membrane subjected to hydration cycling is developed in this paper. The model predicts membrane lifetime as a function of RH cycling amplitude and membrane mechanical properties. The modeling framework consists of three model components: a fuel cell RH distribution model, a hydration/dehydration induced stress model that predicts stress distribution in the membrane, and a damage accrual model that predicts membrane life-time. Short descriptions of the model components along with overall framework are presented in the paper. The model was used for lifetime prediction of a GORE-SELECT membrane.

## I: Introduction:

Mechanical degradation of PEM membrane can limit stack lifetime. Operation of a fuel cell under realistic load cycles results in both chemical and mechanical degradation of the polymer electrolyte membrane. Degradation of the membrane causes opening up of pinholes or crazing of polymer [1, 2, 7] increasing gas-crossover and subsequently resulting in catastrophic failures of the fuel cell stack.

Understanding and modeling the mechanical degradation mechanism and kinetics enables prediction of membrane lifetime as a function of PEM operational conditions and optimization of membrane structure through the choice of reinforcement [4], the membrane processing methods and the operating conditions. The physics based model of membrane mechanical degradation could guide the synthesis of new membranes with tuned mechanical properties and enhanced life in a fuel cell.

Hydration cycling is a primary cause of the mechanical degradation of a geometrically constrained polymer, which exhibits dimensional changes with varied water content. A simple way to impose mechanical damage to a geometrically constrained PEM membrane is to subject it to humidity cycling. At high RH, the membrane absorbs water, and at low RH, the membrane desorbs water. Such RH cycling would result in swelling and shrinking of an unconstrained polymer. RH cycling of a constrained polymer causes cyclic stress. In PEM, such geometrical constraints are imposed by bipolar plate ribs through the gas diffusion layers (GDLs), catalyst layers adjacent to the membrane and at the seals. Moreover, we hypothesize that internal stress in the membrane can be caused by the difference in gas RH at the anode and cathode membrane surfaces that routinely occurs at fuel cell operational conditions. Cyclic stress in the membrane causes irreversible elongation of the membrane [7] and subsequent formation of crazes and cracks that causes gas crossover through the membrane and stack failure.

The goal of this work is to develop a model that predicts membrane lifetime as a function of fuel cell design and operating conditions, and also of membrane transport and mechanical properties. Three components are needed to model the mechanical degradation process under RH cycling. These components are: a fuel cell RH distribution model, a hydration/dehydration induced stress model, and a damage accrual model, see Fig.1. The fuel cell RH distribution model calculates RH distribution in fuel cell gas channels as a function of operating conditions and time. This distribution depends on fuel cell design. The stress model predicts the stress profile in the membrane for a given time-dependent profile of RH at the membrane/electrode interfaces. The damage accrual model predicts the membrane lifetime for a given stress profile in the membrane. The latter two model components use the extended Eyring model of polymer viscoelastic deformation. The damage accrual model component predicts the membrane irreversible elongation as a function of applied stress and time. RH and temperature distribution in gas channels and membrane hydration at steady-state conditions as a function of fuel cell operating conditions were published and discussed in [8, 9, 10, 11, 12]. Hence, they are not described further in this work, and the focus is on other model components. The main features and limitations of the three novel modeling components are briefly discussed below.

**Viscoelastic polymer deformation**

A large number of models of polymer deformation are available in the literature. The linear theory of viscoelasticity was introduced by Boltzmann [13] many decades ago, and it provides the basis for all well-known constitutive models of linear viscoelasticity (Maxwell model, Kelvin–Voigt model, Standard linear solid model, and their generalizations). These models utilize the analogy with mechanical systems consisting of elastic springs and viscous dampers. Models of linear viscoelasticity often fail when either high deformation (>10%) or long-term behavior of polymers is studied. Therefore, to predict membrane lifetime, an approach is required which accounts for non-linear effects of polymer viscoelasticity ..

There are several approaches for non-linear viscoelasticity modeling. The empirical approach uses correlations between time-dependent stress and strain [14, 15 and 16]. This approach relies on numerous fitting parameters that are specific for a given polymer. The alternative approach is a semi-empirical [17, 18] or purely mathematical [19] generalization of Boltzmann's linear theory. The nonlinear, time-dependent constitutive model for prediction of the hygro-thermomechanical behavior of Nafion was proposed in [20] and fitted to experimental data. Neither approach accounts for the microscopic physics of polymer deformation. In contrast to the empirical and semi-empirical models, the molecular theory of non-linear viscoelasticity proposed by Eyring et al [34] is based on a physical concept of polymer dynamics. This model assumes that polymer deformation occurs as a motion of polymer chain segments that overcome potential barriers at the entanglement points. This model predicts the (non-linear) dependence for polymer elongation on applied constant stress. In this paper, the Eyring concept is extended to predict stress relaxation in a constrained polymer.

**Hydration/dehydration stress model**

The hydration/dehydration stress model calculates water distribution in the membrane as a function of time and local stress in the membrane caused by changes in water content. The equilibrium water content of Nafion membranes as a function of gas RH was experimentally studied in [26]. The dynamics of water sorption/desorption by Nafion membranes and water transport through membranes were studied in [27, 28, 29]. Nafion demonstrates unusual water transport properties. For example, its water sorption time is an order of magnitude larger than its water desorption time [27]. Several mathematical models of water transport in Nafion membranes are available in the literature [30, 31, 32 and 33]. Currently, models of ionomer water transport are based on the diffusion equation, and modeling efforts are focused on calculation of the water diffusion coefficient as a function of membrane water content. The ionomer/gas interfacial barrier was hypothesized in [33]. The model developed in [33] explains the peculiar kinetics of water sorption/desorption by the kinetics of water evaporation/condensation at the ionomer/gas boundary. In the current paper, we assume thermodynamic equilibrium of water in the membrane with vapor at the membrane interface, taking advantage of the low ratio of water sorption/desorption time constant to the humidity cycling period.

Water transport in the membrane bulk is treated as diffusion, with the water diffusion coefficient dependent on the membrane water content.

The hydration/dehydration stress model assumes that the local stress in the membrane is induced by competition between the membrane's tendency to swell or shrink in response to a change in hydration level and the geometrical constraints that prevent the membrane from swelling or shrinking. Unconstrained PFSA membranes are known to absorb water and undergo dimensional changes [1]. The dimensional changes can be controlled by the choice of reinforcement [4] and the membrane processing methods. Reinforced membrane demonstrates approximately 4x lower dimensional change than that of non-reinforced membrane under the same conditions. Tang et al. experimentally studied mechanical properties and dimensional change of Nafion 111 membrane [1] and reinforced Gore membrane [4] as a function of temperature and RH. According to [1, 4] the membrane dimensional change is proportional to the change of RH in ambient air. Though, the dependence of water content in the membrane on RH [36] is a nonlinear function, we linearized it in the water content interval from 3 to 12 to speed up the calculations. In the hydration/dehydration stress model, we utilize a new non-linear equation for polymer stress/elongation that is predicted by the extended Eyring model. Taking advantage of this stress/elongation equation, along with the linear dependence of the membrane swelling on water content, we calculate the stress distribution induced by cyclic hydration/dehydration. The model parameters are obtained from experimental data presented in [4, 36].

**Damage accrual model**

The damage accrual model is based on experimental data for membrane failure under cyclic stress and on our concept of polymer plastic deformation. Tensile stress caused by membrane dehydration in a constrained Nafion 111 membrane was studied experimentally by Tang et al. [7]. Dependence of stress on dehydration level was measured and the typical stress value was approximately 1MPa for a membrane that was dehydrated from 100% RH to 60% RH. Such stress is approximately an order of magnitude lower than the tensile strength of a Nafion membrane. Tang et al. [7] have experimentally shown that the amplitude of the cyclic stress that caused substantial permanent elongation of Nafion 111 is 1/10 of the Nafion 111 tensile strength. Nafion membrane creep tests under constant load are reported in the work of Majsztrik et al. [21]. In [21], the Nafion creep dynamics was experimentally studied as a function of the temperature and the membrane hydration level. The applied tensile stress was approximately equal to 1.55 MPa. That stress is also by approximately one order of magnitude lower than the tensile strength of Nafion. A strong dependence of Nafion creep rate on the hydration level was observed in [21]. At 8%RH and >60ºC, low water contenthardens the Nafion membrane and decreases the rate of creep. The creep rate of dry Nafion dramatically increases at ~ 90ºC, which corresponds to the temperature of Nafion α-relaxation detected in DMA tests in [22, 23]. At the same temperature, the creep rate of wet Nafion (RH > 8%) increases much more slowly than that of dry Nafion. Authors explain observed Nafion stiffening with RH increase by the increased electrostatic interaction in ionic clusters composed of $SO_3H$ polar groups in the presence of water molecules. To develop the microscopic model that explains experimental data on Nafion creep dynamics, fundamental insights into the mechanism for polymer damage accrual are needed.

Kusoglu [5] showed through simulations that a hydro-thermal loading under fuel cell operating conditions results in compressive stress in the polymer. According to [5], the stress level exceeds the yield strength, causing permanent damage to the polymer membrane. The compressive stress causes extrusion of membrane materials from the compressed areas such as the areas under the seals. Apparently, the membrane extrusion does not cause the loss of membrane integrity. We assume that the tensile stress in membrane is much more damaging for membrane integrity, because it causes formation and propagation of cracks and crazes. In the present work, we focus on generation of through-plane cracks in the membrane that result in gas crossover through the membrane and dramatic performance loss.

In summary, membrane hydration and dehydration changes the local membrane water content, resulting in a stress cycle in the polymer membrane. A correlation between hydration level and stress cycling is needed. Additionally, to predict life of the membrane, a correlation between membrane macroscopic mechanical properties, stress-cycling, and membrane damage accrual needs to be developed. The overall model that predicts membrane lifetime as a function of membrane mechanical properties and fuel cell operating conditions could guide the synthesis of new membranes with tuned mechanical properties and enhanced life in a fuel cell. Development of such a modeling framework to predict the life of a membrane subjected to hydration cycles is the focus of this work.

The rest of the paper is organized as follows. The extended Eyring model of polymer viscoelastic deformation is presented in Section II. This model is used as a basis of the membrane stress and damage accrual models rather than as a standalone framework component. A model that predicts stress in the membrane during RH cycling in gas channels for given membrane mechanical properties is proposed in Section III. Next, a mathematical model identifying the functional form that relates the stress cycle to irreversible damage of the polymer is discussed in Section IV. The accrual of such irreversible damage results in subsequent polymer failure, determining the membrane life subjected to hydration cycling. Input parameters of the model are summarized in Section V. Modeling results for a GORE-SELECT membrane and discussion of the results are presented in Section VI. Conclusions are presented in Section VII.

## II: Extended Eyring model for polymer deformation

The time dependent response of the stress to applied strain that takes into account viscoelastic relaxation in the polymer material is needed for two components of the degradation model. The model of polymer viscoelastic deformation utilized this paper is based on the assumption that polymer deformation occurs through the transport of polymer chains through entanglements. The early Eyring model [34] is based on a similar concept. The Eyring model calculates the non-linear dependence of polymer deformation on applied constant stress, and is relevant to creep experiments. In a creep experiment, a sample is subjected to constant stress lower than the yield stress, while the elongation is monitored. In this section, we briefly describe a new, extended Eyring model that calculates stress relaxation in a stretched constrained polymer under arbitrary time-dependent controlled deformation.

The concept used in the current model is summarized below. Following [37], we model a polymer as a set of entangled chains. One chain wraps around another chain and turns around at the entanglement point. In the current model, stress is transmitted from one chain to another through the entanglements, which secure the mechanical integrity of the polymer. A Chain Segment (CS) is defined as a fraction of the chain confined between two subsequent entanglements. The macroscopic deformation of the polymer is calculated from the microscopic elongation of CSs driven by the changing distance between entanglements. The macroscopic stress is calculated from the microscopic tension acting along the CSs. This concept to a large extent is similar to the concept of interconnected elastic springs in elasticity theory. The CS is an analog of the spring; the entanglement is an analog of the point where the springs are attached to each other. The elastic deformation in the current polymer model is achieved through the elastic deformation of individual CSs, which is an analog of the elastic deformation of the springs in elasticity theory. The fundamental difference comes from the fact that the chains can slip through entanglements, changing the equilibrium length of CSs. We assume that slippage of the chain through entanglements causes irreversible elongation of the polymer.

To calculate the slip rate as a function of tensile force we consider the dynamics of CS at the entanglement, as indicated in Fig. 2a. In the absence of the tensile force, F, the monomer adjacent to the entanglement is located in the local energy minimum between two symmetric energy barriers $U_0$ (Fig. 2b). The chain slips through the entanglement when the monomer overcomes one of the energy barriers.

In a stressed polymer, a tensile force, F, is induced by stress and acts on the chain (Fig. 2a), disturbing the symmetry of the monomer's potential energy well: the right barrier becomes lower by $Fa$ and the left barrier becomes higher by $Fa$, where $a$ is the monomer length. This results in preferable displacement of the monomer to the right, i.e. along the force direction. Here we assume that the monomer motion is thermally induced and assisted by the force F, $Fa \ll U_0$. The rate $V^+$ at which the monomer jumps to the right is

$$V^+ = \frac{a\omega_0}{2}\exp\left(-\frac{U_0 - Fa}{kT}\right) \quad (1)$$

The rate of monomer displacement to the left $V^-$ is

$$V^- = \frac{a\omega_0}{2}\exp\left(-\frac{U_0 + Fa}{kT}\right) \quad (2)$$

Therefore, the average rate of polymer transport through the entanglement is

$$V = a\omega\sinh\left(\frac{Fa}{kT}\right) \quad (3)$$

Here ω is the frequency of thermally activated jumps of a monomer in the local minimum in the entangled state:

$$\omega = \omega_0 \exp\left(-\frac{U_0}{kT}\right) \tag{4}$$

Equation (3) indicates non-linear dependence of the chain slip rate through the entanglement on the tensile force of the chain. Equation (3) describes the conventional linear dependence of velocity on the force, when $\frac{Fa}{kT} \ll 1$:

$$V = \frac{a^2\omega}{kT} F \tag{5}$$

Here $\frac{a^2\omega}{kT}$ is a mobility of the chain in the entanglement. In the opposite limit, $\frac{Fa}{kT} \gg 1$, equation (3) describes an exponential dependence of velocity on the force:

$$V = \frac{a\omega}{2} \exp\left(\frac{Fa}{kT}\right) \tag{6}$$

In a future publication, the exponential increase of the chain slip rate through entanglements with increasing tensile force will be related to the shape of σ-ε curves observed for polymer materials. We speculate that the yield stress is related to crossover of the chain slip rate from equation (5) to equation (6).

Following the above concept of polymer plastic deformation, we express macroscopic stress, σ, through microscopic tension force, F, and the rate of macroscopic deformation, $\frac{d\varepsilon}{dt}$, through the microscopic chain velocity, V, predicted by equation (3). The detailed derivation will be published in a forthcoming paper. The final microscopic equation that relates polymer deformation, ε, with external stress, σ, is presented below

$$\frac{d\sigma}{dt} = E\frac{d\varepsilon}{dt} - \frac{\sigma_T}{\tau}\sinh\left(\frac{\sigma}{\sigma_T}\right) \tag{7}$$

Here τ is the polymer relaxation time and $E = \frac{\partial \sigma}{\partial \varepsilon}$ is Young's modulus for the polymer.

$\sigma_T = \frac{kT}{V_{rel}}$, where $V_{rel}$ is a typical volume of polymer matrix around the entanglement disturbed by one elementary act of monomer transfer through the entanglement. The first term in right hand part of equation (7) is the rate of the stress change induced by polymer deformation. The second term is rate the plastic relaxation of the stress. This term represents an extension of Maxwell's relaxation rate, σ/τ. In the case of low stress, $\sigma \ll \sigma_T$, hyperbolic sine, sinh, in the right hand side of equation (7) is expanded to reproduce the well-known Maxwell equation for polymer viscoelastic deformation:

$$\frac{\partial}{\partial t}[\sigma - E\varepsilon] = -\frac{\sigma}{\tau} \qquad (8)$$

The fundamental difference between equation (7) and the Maxwell equation (8) is in the relaxation terms in the right hand sides of equations (7) and (8). At large stress $\sigma \gg \sigma_T$ the hyperbolic sine asymptotically approaches the exponent, leading to a sharp increase in the stress relaxation rate as the applied stress is increased. The detailed derivation of equation (7) and validation of the model presented above will be published later in a separate paper.

## III: Membrane Stress Model

In this section, we present the model that calculates the stress induced by hydration/dehydration cycling. In a fuel cell stack, the membrane electrode assembly (MEA) is constrained between two bipolar plates (Fig. 3a). The membrane can neitherbend nor change length. Fuel is supplied to the MEA through fuel gas channels and oxidant is supplied through oxidant (air) channels as indicated schematically in Figure 3.

To illustrate the mechanism of membrane failure due to RH cycling in the gas channels we consider the following experiment. Initially, wet fuel is fed into the anode gas channel (top channel shown in Figure 3a) and wet air is fed into the cathode gas channel (bottom channel shown in Figure 3a). Both sides of the membrane are equilibrated with wet gas until the equilibrium water content $\lambda_0$ is reached and all mechanical stresses in the membrane relax to zero. The cathode side of the membrane dehydrates and attempts to shrink when feed gas (air) is switched from wet to dry in the cathode gas channel. Mechanical constraints which prevent the membrane from shrinking, cause in-plane tensile stress, $\sigma_{yy}$, in the membrane. Tensile stress results in a crack formation on the RH cycled side of the membrane (Fig. 3b). Further propagation of the crack in the through-plane direction causes membrane mechanical failure.

To calculate water distribution in the membrane cross-section at a fixed coordinate y along the air channel we use RH(y,t), the relative humidity distribution function predicted by the fuel cell model. Here we assume that the equilibration time of water in the membrane is much smaller than RH cycling period in the gas channels. Therefore, the water at the membrane interfaces is in thermodynamic equilibrium with vapor in the gas channels:

$$\lambda(x=0,y,t) = \lambda_{eq}(RH_C(y,t))$$
$$\lambda(x=L_m,y,t) = \lambda_{eq}(RH_A(y,t))$$
(9)

Here x is the through-membrane coordinate, $\lambda_{eq}(RH)$ is the equilibrium water content in the membrane (an experimentally measured function of RH), $RH_C$ is the gas RH in the cathode gas channel, and $RH_A$ is the gas RH in the anode gas channel. In the current paper, we utilize the following conventional diffusion equation, with a water content dependent diffusion coefficient, to calculate water profile in the membrane:

$$\frac{\partial \lambda}{\partial t} - \frac{\partial}{\partial x}\left(D(\lambda)\frac{\partial \lambda}{\partial x}\right) = 0 \tag{10}$$

Here $D(\lambda)$ is the experimentally measured water diffusion coefficient in the membrane (see Fig.7 in Section VI). Initial conditions for equation (10) can be chosen arbitrarily, because after several cycles the membrane evolves to a new quasi-equilibrium state governed by cyclic conditions, and the initial state becomes irrelevant. We chose the initial conditions $\lambda(x,0) = \lambda_0$. Solution of equation (10) with the boundary conditions (9) gives the water distribution in the membrane, $\lambda(x,y,t)$. Below, we calculate the membrane stress at the membrane boundary adjacent to the air inlet (fixed coordinate y) where RH cycling and stress amplitudes are maximal and limit membrane lifetime. We start with a calculation of the linear elastic response of the membrane to small changes in $\lambda$ and subsequently model the more general viscoelastic case.

The macroscopic state of the membrane is determined by two parameters: the membrane length, $L_0$, and water content, $\lambda_0$. We assume that at equilibrium, the stress in the membrane is equal to zero. The stress in the membrane can be generated by deviation of the membrane length from equilibrium, $\Delta L = L - L_0$, at constant water content. Also, the stress can be generated by deviation of water content from equilibrium, $\Delta\lambda(x,t) = \lambda(x,t) - \lambda_0$, in the geometrically constrained membrane (at constant length). By analogy with linear elasticity theory with thermal expansion, we calculate the linear response of the membrane stress to small deviation from equilibrium as follows:

$$\sigma_{yy} = E\left(\varepsilon_{yy} - \alpha\Delta\lambda(x,t)\right) \tag{11}$$

Here $\varepsilon_{yy} = \Delta L/L_0$ and E is the Young modulus of the membrane. Swelling coefficient α determines the length change, $\Delta y$, of unconstrained (zero σ) membrane caused by the change of the membrane water content $\Delta\lambda$: $\Delta y/y = \alpha\Delta\lambda$. In 1D approximation, the local stress in the membrane depends on the change of the local membrane hydration, (λ(x,t)-λ$_0$).

In the membrane, stress can be induced by changes in deformation, water content and simultaneous plastic relaxation. The total rate of the stress change is the sum of the stress change rate caused by membrane deformation and the stress change rate caused by the change of membrane water content

$$\frac{d\sigma_{yy}}{dt} = \left.\frac{\partial\sigma_{yy}}{\partial t}\right|_\varepsilon + \left.\frac{\partial\sigma_{yy}}{\partial t}\right|_\lambda \tag{12}$$

According to equation (11) the rate of the stress change induced by the change of the rate of deviation of water content in constrained system with constant deformation and without plastic relaxation is

$$\left.\frac{\partial\sigma_{yy}}{\partial t}\right|_\varepsilon = E\alpha\frac{d\Delta\lambda}{dt} \tag{12a}$$

The second term in equation (12) is presented by equation (7)

$$\left.\frac{\partial \sigma_{yy}}{\partial t}\right|_\lambda = E\left.\frac{d\varepsilon_{yy}}{dt}\right|_\lambda - \frac{\sigma_T}{\tau}\sinh\left(\frac{\sigma}{\sigma_T}\right) \tag{12b}$$

Substituting equations (12a) and (12b) into the right hand side of equation (12) we obtain the equation for the total rate of stress change in the membrane

$$\frac{d\sigma_{yy}(x,t)}{dt} = E\left.\frac{\partial \varepsilon_{yy}(x,t)}{\partial t}\right|_\lambda + E\alpha\frac{d\Delta\lambda(x,t)}{dt} - \frac{\sigma_T}{\tau}\sinh\left(\frac{\sigma_{yy}(x,t)}{\sigma_T}\right) \tag{13}$$

The first term in the right hand side of equation (13) is equal to zero in a constrained membrane. Local water content, λ(x,t), is calculated by equation (10) with boundary conditions (9). Equation (13) relates the stress in the membrane with water content in the membrane for arbitrary RH cycling protocol.

Equation (13) was solved numerically with parameters specified in the Table 1. The calculated stress as a function of time is presented in Figure 7. The results indicate that the membrane stress becomes a periodic function of time after several cycles if RH is a periodic function of time. To derive an analytical equation for the periodic stress we utilize the separation of slow variable method. The plastic (irreversible) deformation of the membrane during one cycle is small because the cycling period is much smaller than the membrane relaxation time, $T_{cyc} \ll \tau$. Therefore, we can consider the membrane as an elastic media during one cycle. However, the membrane slowly approaches a new equilibrium state during each cycle. After a long time, $t \gg \tau$, the membrane reaches the new state with new equilibrium water content, $\overline{\lambda}(x)$. During the cycle, λ oscillates below and above $\overline{\lambda}(x)$. At $\lambda = \overline{\lambda}(x)$ the membrane is not stressed. The in-plane stress in the periodical regime is driven by the deviation of λ from $\bar{\lambda}(x)$ and is calculated by following equation:

$$\sigma_{yy}(x,t) = -E(x,t)\alpha(\lambda(x,t) - \bar{\lambda}(x)) \qquad (14)$$

The stress averaged over time in the periodic regime is equal to zero for any x. Averaging (14) over time and taking advantage of the condition $\langle \sigma_{yy}(x,t) \rangle_t = 0$, we obtain the following expression for $\bar{\lambda}(x)$:

$$\bar{\lambda}(x) = \frac{\langle E(x,t)\lambda(x,t) \rangle_t}{\langle E(x,t) \rangle_t} \qquad (15)$$

Substituting (15) into (14) we obtain the final equation for the stress under periodic RH cycling:

$$\sigma_{yy}(x,t) = -E(x,t)\alpha\left(\lambda(x,t) - \frac{\langle E(x,t)\lambda(x,t) \rangle_t}{\langle E(x,t) \rangle_t}\right) \qquad (16)$$

Substituting $\lambda(x,t)$ calculated from equation (10) with boundary conditions (9) into equation (16), we calculate cyclic stress in the membrane.

## IV: Damage accrual model

In this section, we describe the model that calculates the membrane lifetime as a function of applied cyclic stress. Using the membrane stress model presented in Section III, we calculate the stress profile in the membrane, $\sigma_{yy}(x,t)$, for a given water content, $\lambda(x,t)$. We predict the membrane lifetime as a function of membrane stress using the damage accrual model.

The typical stress in the membrane under fuel cell operating conditions is much smaller than the membrane tensile strength. Many RH cycles are required to cause substantial membrane damage. However, with each cycle, the cyclic stress causes small irreversible elongation, i.e. plastic deformation, of the membrane. Accumulation of irreversible elongation in the membrane causes membrane damage after a large number of cycles.

Tang et al. [7] experimentally studied Nafion111 membrane irreversible elongation under cyclic stress. They observed that irreversible elongation is accumulated over a large number of cycles and causes damage of the membrane even at relatively small amplitudes of cyclic stress. They also demonstrated that at relatively small stress, $\sigma<4MPa$, the membrane elongation rate is a linear function of applied stress magnitude. At larger stress amplitudes of about $\sigma=6.5MPa$, the elongation rate rapidly increases. We speculate that such nonlinear dependence of elongation rate on stress amplitude is caused by exponential dependence of the stress relaxation rate on the stress magnitude predicted by extended Eyring model presented in Section III.

The elongation of the membrane subjected to step-like cyclic stress is a function of the stress amplitude $\sigma$ and the cycle period $T_{cyc}$. Applying equation (7) to the membrane elongation during constant stress hold in one cycle we obtain:

$$\varepsilon_1 = \frac{\sigma_T T_{cyc}}{E\tau} \sinh\left(\frac{\sigma}{\sigma_T}\right) \qquad (17)$$

Only a small fraction, $\gamma$, of this elongation is irreversible, i.e. leads to membrane damage. In the current work we assume that $\gamma$ depends only on cyclic period, $T_{cyc}$, and is independent of $\sigma$. Also, we do not account for self-acceleration of membrane damage, i.e. we assume that irreversible elongation is proportional to the number of cycles N. The equation for irreversible elongation of the membrane per N cycles is:

$$\varepsilon(\sigma, N) = N\varepsilon_0 \sinh\left(\frac{\sigma}{\sigma_T}\right) \qquad (18)$$

Here $\varepsilon_0$ is:

$$\varepsilon_0 = \frac{\sigma_T T_{cyc} \gamma(T_{cyc})}{E\tau} \qquad (19)$$

The elongation as a function of the stress and the number of cycles, $\varepsilon(\sigma,N)$ was measured experimentally using Dynamic Mechanical Analysis (DMA) equipment for two specific values of stress amplitude. Because $\varepsilon_0$ is proportional to an unknown parameter $\gamma$ and inversely proportional to another unknown parameter $\tau$, the lumped parameter $\varepsilon_0$ was used for fitting. The model parameters $\varepsilon_0$ and $\sigma_T$ were fitted to DMA data. The membrane elongation for arbitrary stress amplitude $\sigma$ was calculated from equation (18).

To validate equation (18), elongation per one stress cycle, $\varepsilon_1(\sigma)$, was calculated from Tang's experimental data [7] for Nafion111 membrane and plotted as a function of applied stress, $\sigma$, see Fig. 4. It follows from (18) that

$$\varepsilon_1(\sigma) = \frac{d\varepsilon(\sigma, N)}{dN} = \varepsilon_0 \sinh\left(\frac{\sigma}{\sigma_T}\right) \qquad (20)$$

The experimental data are in good agreement with the prediction of equation (20) with $\sigma_T=1$ MPa and $\varepsilon_0=3.7\cdot10^{-6}$ (Fig. 4).

Membrane failure occurs after a critical irreversible elongation (damage), $\varepsilon_{crit}$, is accumulated. The number of cycles to failure, $N_{crit}$, is calculated from the following condition:

$$N_{crit}\varepsilon_1(\sigma) = \varepsilon_{crit} \quad (21)$$

We assume that $\varepsilon_{crit}$ does not depend on stress amplitude and frequency and can be measured once at a specific stress in a DMA test. The number of cycles to failure, $N_{DMA}$, in out-off-cell tests is calculated by equation

$$N_{crit} = \frac{N_0}{\sinh(\sigma/\sigma_T)} \quad (22)$$

$$N_0 = \frac{\varepsilon_{crit}}{\varepsilon_0} = \frac{\varepsilon_{crit} E \tau}{\sigma_T T_{cyc} \gamma(T_{cyc})} \quad (23)$$

We use $N_0$ as a lumped fitting parameter obtained from a DMA test. To obtain the values of $N_0$ and $\sigma_T$ we performed stress cycling for two stress amplitudes, $\sigma$, and fitted the experimental number of cycles to failure, $N_{crit}$, by equation (22). Equation (22) predicts membrane lifetime in a DMA test with arbitrary stress amplitude $\sigma$ using the obtained values of $N_0$ and $\sigma_T$ from experimental data. Analytical dependence of these parameters on temperature and RH will be published later.

Fuel cell conditions under RH cycling differ from DMA test conditions because in DMA testing the stress is uniformly distributed through the membrane cross-section, while in the cell, the stress in the membrane changes substantially from the anode to the cathode side. We incorporated calibration parameter $\beta$ into the current version of the model in order to use equation (22) for in-cell membrane lifetime prediction. This parameter is used for renormalization of stress. The membrane damage in DMA testing with stress amplitude σ is equivalent to membrane damage in a fuel cell stack when the membrane is subjected to RH cycling that cause a stress amplitude of β·σ.

To obtain the calibration parameter β we performed a single RH cycling experiment in a fuel cell up to membrane failure for particular conditions and measured the experimental membrane lifetime $N_{cell}$. Using the membrane stress model we calculated the stress amplitude $\sigma_{max}$ for these particular conditions. Parameter β is calculated from the following equation:

$$N_{cell} = \frac{N_0}{\sinh(\beta\sigma_{max}/\sigma_T)} \quad (24)$$

Using the parameters $N_0$ and $\sigma_T$ obtained from fitting DMA data by equation (22), and the parameter β obtained from the in-cell test, we predict the in-cell membrane lifetime under arbitrary RH cycling conditions.

## V: Model parameters

In this section we present input model parameter values and experimental data for a GORE-SELECT® membrane in a PRIMEA® MEA. In RH cycling experiments, hydrogen with 100% RH was fed to the anode channel. Air fed to the cathode channel was cycled from $RH_{max}$ to $RH_{min}$, with equal time intervals. Experimental conditions are summarized in Table 1. Mechanical properties of GORE-SELECT membranes were obtained from available literature and summarized in Table 1. DMA tests required for model calibration were also performed.

# VI: Model results

In this section we report the result of modeling of GORE-SELECT membrane damage and lifetime under in-cell RH cycling conditions. The model components, i.e. fuel cell model, membrane stress model and stress accrual model were incorporated into the MATLAB code. The model parameters and test conditions are summarized in Section V.

The relative humidity variation in the cathode gas channel during the cycle was calculated from the fuel cell model and used as the input for the membrane stress model. The calculated relative humidity variation is shown in Fig.6. The function with fast exponential relaxation was used instead of step-wise function for better convergence of the numerical solution. The calculated dependence of $\lambda$ on time at three points across the membrane (at the cathode side, at the middle of the membrane and at the anode side) is also shown in Fig.6. This dependence is calculated from water diffusion equation in the membrane (10) with boundary conditions (9) and initial condition $\lambda(x,0) = 11$. Figure 9 indicates that the membrane water content λ(t) follows the humidity cycle RH(t). There is no time lag because water diffusion time in the membrane $t_{dif} = \frac{h^2}{D} \approx 0.2s$ is much smaller than cycling period $T_{cyc}$=30s. Here h=18 μm is membrane thickness and D=0.5*10$^{-6}$ cm$^2$/s is the average water diffusion coefficient in the membrane.

Water content in the membrane, $\lambda(x,t)$, is utilized in the membrane stress model to predict the stress $\sigma(x,t)$ distribution in the membrane. The model prediction of the stress in the membrane by equation (13) as a function of time is shown in Fig.7 at the cathode and anode sides of the membrane. The membrane reaches quasi-steady state regime after t >> τ, and stress becomes a periodic function of time, oscillating around the equilibrium value. The model prediction for the stress in the periodic regime as a function of time calculated by equation (16) is shown in Fig.8. Maximal stress is imposed in the region with maximal humidity variation, i.e. at the cathode side of the membrane. At the anode side of the membrane the stress is equal to zero because there is no humidity cycling. One can see that the stress calculated by equation (13) at large time is in good agreement with the periodic stress shown in Fig.8, calculated analytically from equation (16).

Figures 7 and 8 indicate that after time t larger than relaxation time τ the stress at the cathode sides oscillates around the equilibrium stress $\sigma_{eq} = 0$. The membrane at the cathode side is subjected to tensile stress when $\lambda(0,t) < \bar{\lambda}(x=0)$, and to compressive stress when $\lambda(0,t) > \bar{\lambda}(x=0)$.

The maximal value of the stress calculated by the membrane stress model is used in the damage accrual model to predict membrane lifetime by equation (24). As discussed in Section IV, this equation contains three unknown parameters, $\sigma_T$, $N_0$ and β. Parameters $\sigma_T$ and $N_0$ are calculated by fitting UTC POWER DMA data by equation (22). Calibration parameter β is calculated from an in-cell lifetime experiment under specific RH cycling conditions. The in-cell membrane stress was calculated by the membrane stress model.

Using the model parameters calculated from DMA testing and in-cell RH cycling, we predicted GORE-SELECT membrane lifetime under arbitrary RH cycling conditions. Predicted membrane lifetime (number of cycles to failure) as a function of minimal RH in the cathode gas channel, $RH_{min}$, is plotted in Fig. 9. The anode gas humidity was assumed to be 100%. The maximal RH in the cathode gas channel, $RH_{max}$, of 100% was also assumed. At high RH cycling amplitude (low $RH_{min}$) the model predicts exponential decrease of membrane lifetime with increase of RH cycling amplitude. Stress amplitude is approximately proportional to RH cycling amplitude, and the number of cycles to failure depends exponentially on the stress for large stress according to equation (22). This results in exponential dependence of the membrane lifetime on RH cycling amplitude for large values of amplitude (small value of $RH_{min}$).

## VII: Conclusions

A modeling framework was developed that predicts the lifetime of PEM fuel cell membranes subjected to hydration cycling. The developed model predicts membrane lifetime as a function of RH cycling amplitude and membrane mechanical properties. Membrane failure in the fuel cell is caused by damage accumulation under cyclic stress in the membrane subjected to cyclic hydration/dehydration. A fuel cell membrane is typically subjected to hydration/dehydration under fuel cell conditions. One side of the membrane dehydrates and attempts to shrink when the feed gas (air) is switched from wet to dry in the cathode gas channel. Mechanical constraints imposed by bipolar plates prevent the membrane from shrinking. This causes in-plane tensile stress in the membrane. Tensile stress causes a crack formation at RH cycled side of the membrane (Fig. 3). Further propagation of the crack in the through-plane direction causes membrane mechanical failure.

The modeling framework consists of three components: a model of RH distribution in gas channels, a membrane stress model, and a damage accrual model. Several models of RH distribution in gas channels are available in literature [8, 9, 10, 11 and 12] and we do not discuss this model component in the current paper. In the current version of the model, we assume equilibrium of water in the membrane with the vapor at the membrane/gas interface and use the conventional diffusion equation for calculation of water content in the membrane bulk, with the water diffusion coefficient dependent on water content, $D(\lambda)$. The membrane stress model calculates the stress in the membrane caused by membrane cyclic swelling/shrinking under RH cycling conditions. The local stress in the membrane is caused by the change in the local hydration level. The damage accrual model predicts the number of cycles to failure for the membrane under applied cyclic stress. The input for the damage accrual model is a maximal stress in the membrane calculated by the membrane stress model and experimental membrane lifetimes in DMA tests for two cycling amplitudes. The current version of the model also contains one calibration parameter obtained from an in-cell RH cycling experiment with specific RH cycling conditions.

The model was utilized for in-cell lifetime predictions of GORE-SELECT membranes. The membrane mechanical properties and swelling coefficient were obtained from the literature. DMA testing and in-cell RH cycling were carried out at UTC Power. After calibration, the model predicts membrane lifetime in fuel cells under arbitrary RH cycling conditions. The calculated membrane lifetime (number of cycles to failure) as a function of minimal RH in cathode gas channel, $RH_{min}$, is plotted in Fig. 9. At high RH cycling amplitude (low $RH_{min}$) the model predicts exponential decrease of membrane lifetime with increase of RH cycling amplitude. Stress amplitude is approximately proportional to RH cycling amplitude, and the number of cycles to failure depends exponentially on stress for large stress magnitude according to equation (22). This results in exponential dependence of membrane lifetime on RH cycling amplitude for large values of amplitude (small value of $RH_{min}$).

## Acknowledgement

This work was supported by the US Department of Energy under the grant DE-PS36-08GO98009.

## List of Tables

Table 1  Model parameters and experimental conditions

| Parameter | Description | Value/Source |
|---|---|---|
| **Operational conditions** | | |
| $T_{cyc}$ | RH cycling time period | 30 sec |
| $RH_{max}$ | Maximum relative humidity in cathode channel | 100% |
| $RH_{min}$ | Minimum relative humidity in cathode channel | 10% |
| T | Cell temperature | 78ºC |
| **Membrane properties** | | |
| E(RH,T) | Young's modulus of the membrane | [4] |
| $\lambda(a)$ | Membrane water content as a function of vapor activity [mol H$_2$O/mol SO$_3$] | [36] |
| $D(\lambda)$, | Diffusion coefficient of water in the membrane | Fig.5 |

| $\alpha$ | Dimension change coefficient | 0.004 (data from [4]) |
| $\sigma_T$ | Fitting parameter (internal data fit to Eq. 27) | 1.7 MPa |
| $N_0$ | Fitting parameter (internal data fit to Eq. 27) | $1.2 \cdot 10^6$ |
| $\beta$ | DMA to in-cell calibration parameter | 2.3 |

# Figure Captions

Figure 1. Schematic diagram showing the components of the model framework to predict mechanical life of the membrane undergoing hydration cycles.

Figure 2 (a) Polymer chain segment constrained by one entanglement and subjected to the external force F. (b) Potential energy of monomer in entanglement before applying the force (in the left hand side of the Fig. 2b) and after applying the force (in the right hand side of the Fig. 2b).

Figure 3. Schematic figure of membrane in fuel cell (counter-flow configuration) constrained between two bi-polar plates (a) and crack formation (b). The wet fuel exits from the anode gas channel (top channel). Initially, wet air is fed into cathode gas channel (bottom channel). When cathode side feed gas (air) is switched from wet to dry, cathode side of the membrane dehydrates and attempts to shrink. The anode side of the membrane remains wet and swollen.

Figure 4. Fitting Tang's experimental data [7] (triangles) by equation (22) (solid line). Fitting parameters values are: $\sigma_T$=1 MPa and $\varepsilon_0$=3.7·10$^{-6}$.

Figure 5. Dependence of diffusion coefficient in Nafion membrane on water content $\lambda$ (UTRC internal experimental data). There are not available data in range from $\lambda = 2.5$ to $\lambda = 15$ so linear approximation is used in this region.

Figure 6. Model prediction of dependence of RH on time and $\lambda$ on time at the anode side, at the middle of the membrane and at the cathode side.

Figure 7. Model prediction of stress as a function of time under RH cycling at the cathode and at the anode sides of membrane.

Figure 8. Model prediction of stress in quasi-steady-state regime at the cathode side, at the middle of the membrane and at the anode side.

Figure 9. Calculated membrane lifetime (the number of cycles to failure) as a function of minimal RH at the cathode side of membrane. Anode side of membrane maintains 100% RH.

# Figures

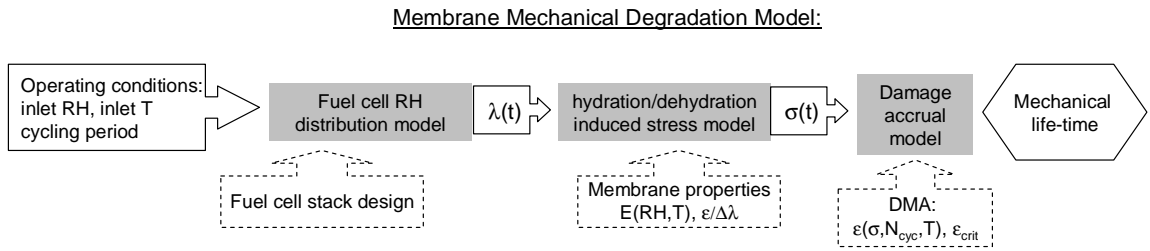

Figure 1

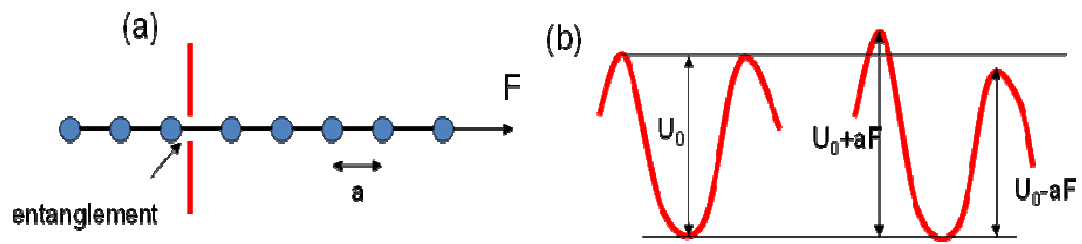

Figure 2

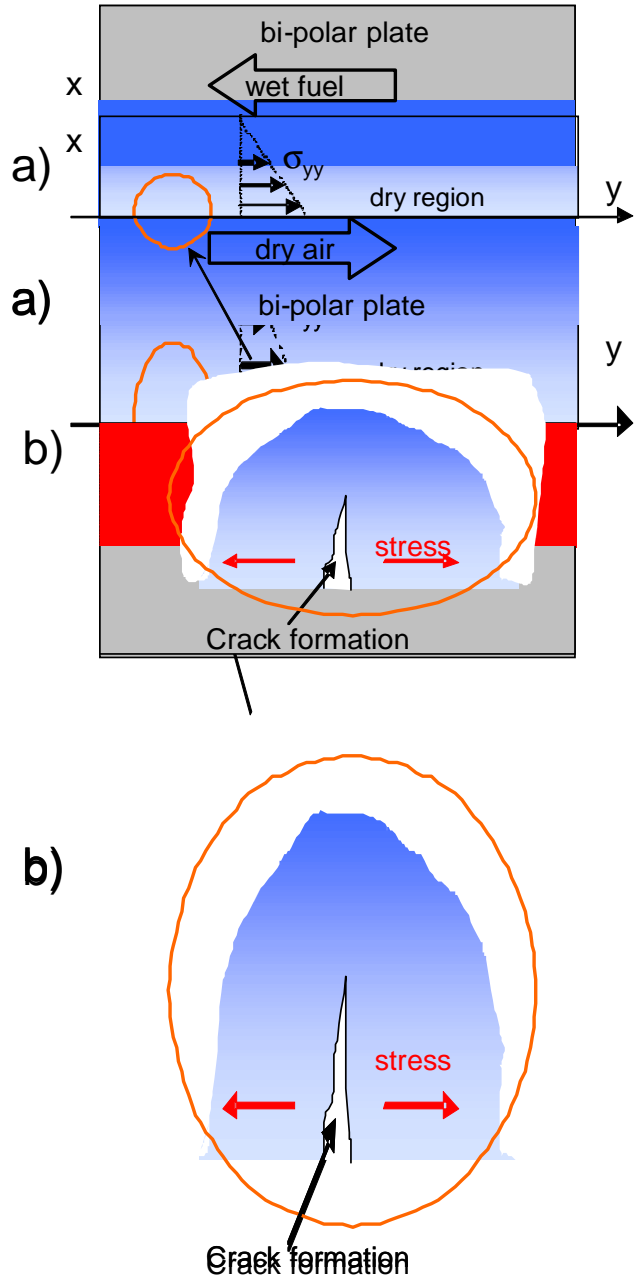

Figure 3

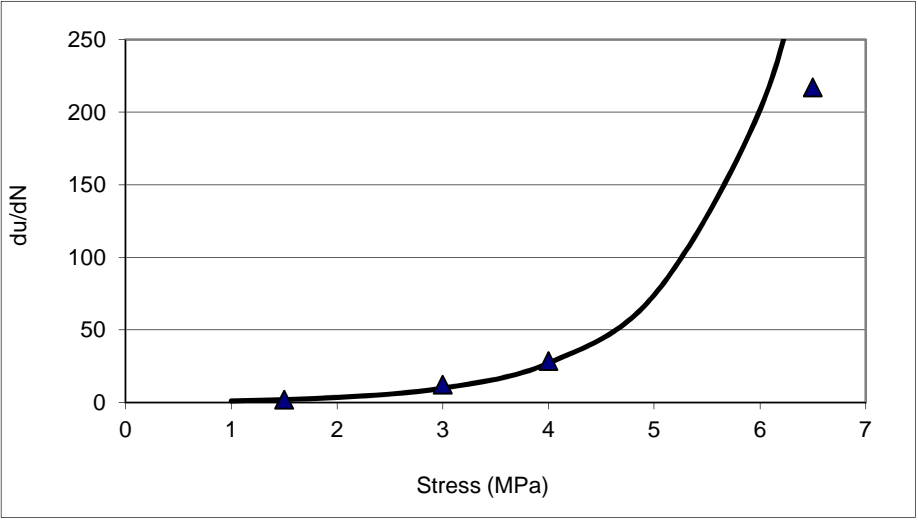

Figure 4

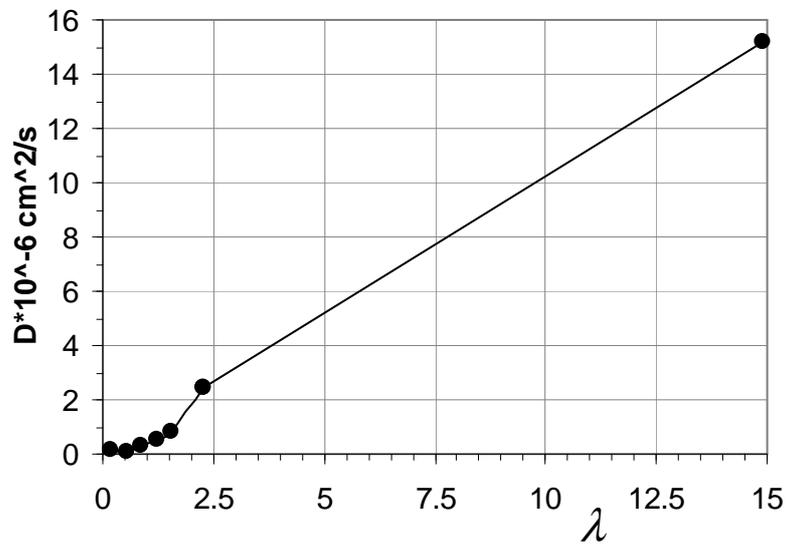

Figure 5

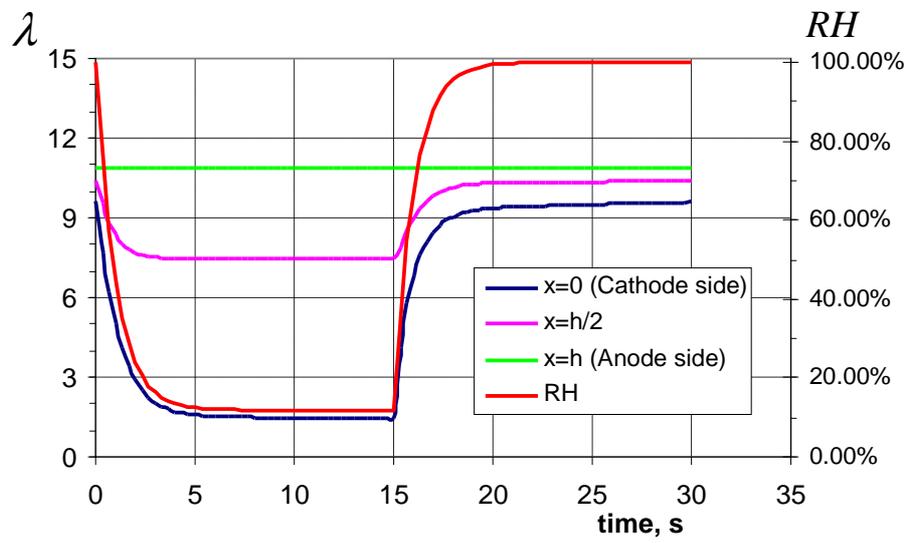

Figure 6

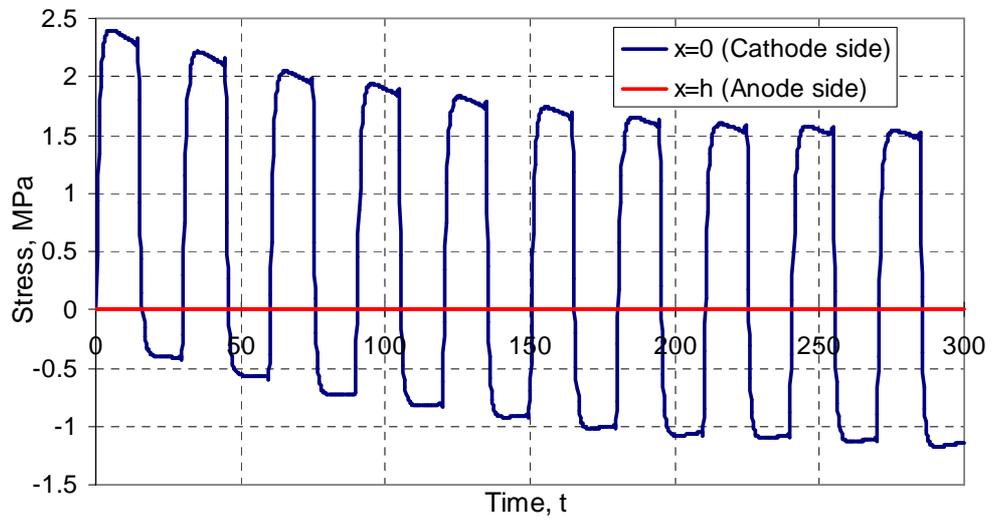

Figure 7

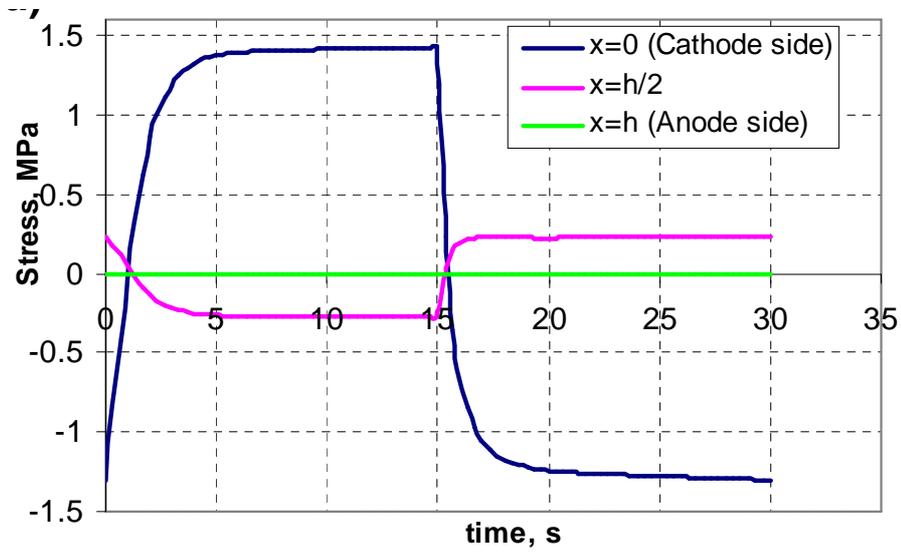

Figure 8

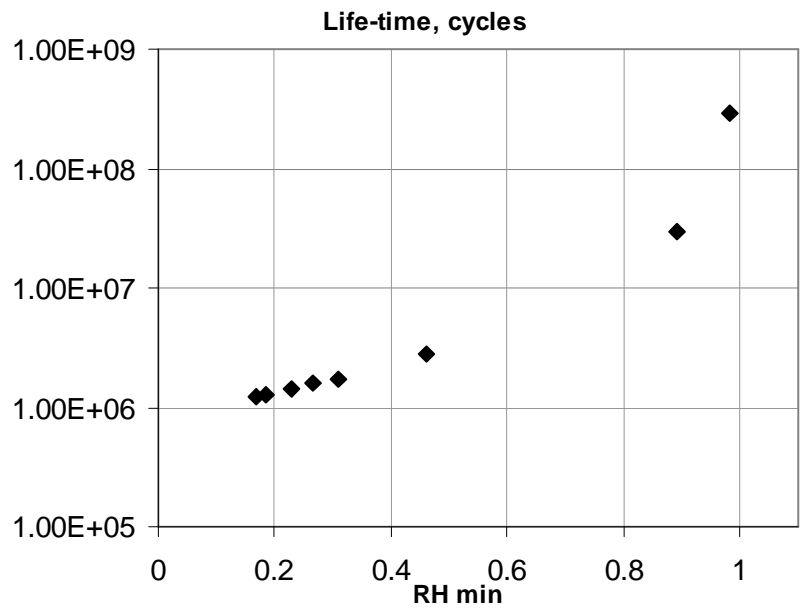

Figure 9